\documentclass[aip, jcp, amsmath, amsthm, amssymb, citeautoscript, reprint]{revtex4-2}
\usepackage{amsmath, amsthm, amssymb}
\usepackage{graphicx, float, xcolor, pgf}
\usepackage{physics, siunitx}
\usepackage[caption=false, justification=centerlast]{subfig}
\usepackage{gensymb}
\usepackage{hyperref}
\hypersetup{
    pdfauthor={Devansh Sharma and Amartya Bose},
    pdftitle={Routes of Transport in the Path Integral Lindblad Dynamics through State-to-State Analysis},
    colorlinks=true,
}
\usepackage[normalem]{ulem}

\allowdisplaybreaks
\begin{document}
\title{Routes of Transport in the Path Integral Lindblad Dynamics through State-to-State Analysis}
\author{Devansh Sharma}
\affiliation{Department of Chemical Sciences, Tata Institute of Fundamental Research, Mumbai 400005, India}
\author{Amartya Bose}
\email{amartya.bose@tifr.res.in}
\email{amartya.bose@gmail.com}
\affiliation{Department of Chemical Sciences, Tata Institute of Fundamental Research, Mumbai 400005, India}
\begin{abstract}
Analyzing routes of transport for open quantum systems with non-equilibrium
initial conditions is extremely challenging. The state-to-state approach~[A.
Bose, and P.L. Walters, J. Chem. Theory Comput. 2023, \textbf{19}, 15,
4828–4836] has proven to be a useful method for understanding transport
mechanisms in quantum systems interacting with dissipative thermal baths, and
has been recently extended to non-Hermitian systems to account for empirical
loss. These non-Hermitian descriptions are, however, not capable of describing
empirical processes of more general nature, including but not limited to a
variety of pumping processes. We extend the state-to-state analysis to account
for Lindbladian descriptions of generic dissipative, pumping and decohering
processes acting on a system which is exchanging energy with a thermal bath.
This Lindblad state-to-state method can elucidate routes of transport  in
systems coupled to a bath and additionally acted upon by Lindblad jump
operators. The method is demonstrated using examples of excitonic aggregates
subject to incoherent pumping and draining processes. Using this new
state-to-state formalism, we demonstrate the establishment of steady-state
excitonic currents across molecular aggregates, yielding a different
first-principles approach to quantifying the same.
\end{abstract}

\maketitle
\section{Introduction}\label{sec:intro}

Transport processes are ubiquitous. Be it the exciton transport in
light-harvesting antenna complexes in photosynthetic
systems~\cite{pulleritsPhotosyntheticLightHarvestingPigmentProtein1996,
wuEfficientEnergyTransfer2012, ishizakiQuantumCoherenceinPhotosyntheticLH2012,
bakerRobustnessEfficiencyOptimality2015} or the charge transport in molecular
wires and solar cells~\cite{ratnerMolecularWires1998,
mottaghianChargeTransportModelSolarCell2013}, transport plays a pivotal role in
their functioning. Simulating and understanding such processes is
instrumental for getting insights into the functioning of various molecular
systems as well as ideas for designing new materials. However, these simulations involving quantum particles in condensed phases are complicated.
Tackling the already large number of electronic degrees of freedom of complex
aggregates, which is what such transport systems generally are, along with a
plethora of environmental modes with limited computational resources and time,
requires some thought. Approximate methods like
Redfield~\cite{blochGeneralizedTheoryRelaxation1957,
redfieldTheoryRelaxationProcesses1957} and
F\"{o}rster~\cite{forsterZwischenmolekulareEnergiewanderungUnd1948} are
popularly used but cannot be applied in non-perturbative regimes. Wave
function-based methods like density matrix renormalization
group~\cite{whiteDensityMatrixFormulation1992,
schollwockDensitymatrixRenormalizationGroup2005} (DMRG) and multi-configuration
time-dependent Hartree~\cite{beckMulticonfigurationTimedependentHartree2000,
wangMultilayerFormulationMulticonfiguration2003} (MCTDH), while capable of
handling non-perturbative environments, are unable to efficiently deal with large
number of environment modes with thermally populated initial
conditions~\cite{tanimuraNumericallyExactApproach2020}.

Methods based on reduced density matrices (RDMs) provide an efficient route to
simulating the dynamics of these thermal systems. Most of them are
based on the Feynman path integral
formalism.~\cite{feynmanQuantumMechanicsPath2010} The quasi-adiabatic propagator
path integral~\cite{makriTensorPropagatorIterativeI1995,
makriTensorPropagatorIterativeII1995} (QuAPI) and hierarchical equations of
motion~\cite{tanimuraQuantumClassicalFokkerPlanck1991,
tanimuraNumericallyExactApproach2020} (HEOM) are two of the most commonly used
families of approaches. Both these methods have historically been notorious for being
computationally expensive. A series of recent
developments~\cite{strathearnEfficientNonMarkovianQuantum2018,
jorgensenExploitingCausalTensor2019, boseTensorNetworkRepresentation2021,
boseMultisiteDecompositionTensor2022, bosePairwiseConnectedTensor2022,
makriSmallMatrixPath2020, makriSmallMatrixModular2021, xuTamingQuantumNoise2022}
have made it possible to apply these methods to simulate larger systems as
well~\cite{boseTensorNetworkPath2022, kunduB800toB850RelaxationExcitation2022}.

However, these numerically exact methods require a proper parameterization of
both the system Hamiltonian and the system-environment interactions in the form
of a spectral density~\cite{caldeiraPathIntegralApproach1983,
makriLinearResponseApproximation1999, boseZerocostCorrectionsInfluence2022}.
This can prove to be challenging for pumping and loss (draining) processes, among others, frequently encountered in transport systems. For instance, in the case of
exciton-polaritons, the Fabry-P\`erot cavities that are involved almost always
have imperfections that lead to a possibility of the loss of a photon. The
proper characterization of the ``bath'' that causes this may often be well-nigh
impossible. Additionally, there are other cases, where these extra interactions
may be characterized, but incorporating them at an exact level can increase the
cost of simulations exponentially. As a result, in most studies of the excitonic
transport dynamics in light-harvesting complexes, the mechanism of ``extraction'' of
the exciton is completely ignored,~\cite{ishizakiUnifiedTreatmentQuantum2009,
ishizakiTheoreticalExaminationQuantum2009, boseAllModeQuantumClassical2020,
kunduB800toB850RelaxationExcitation2022, boseTensorNetworkPath2022} leading to a
rise of population in the ``sink'' site. But we know that these are highly coupled
systems and adding the draining mechanism can change the details of the
transport itself. Therefore, the question is whether it is possible to
incorporate the ``sources'' and ``sinks'' in an empirical fashion through the relevant
time-scales, which may be more easily obtained.

Semiclassical methods have
recently been successfully coupled with empirical terms from the Lindblad master
equation~\cite{lindbladGeneratorsQuantumDynamical1976,
goriniCompletelyPositiveDynamical1976} ensuring the ability to describe the
vibronic degrees of freedom at a semiclassical level and the loss terms
empirically.~\cite{mondalQuantumDynamicsSimulations2023} If one wants to simulate the system in a fully quantum manner, incorporation of non-Hermitian descriptions of the
system in path integrals~\cite{palmNonperturbativeEnvironmentalInfluence2017} provides a simple approach. While this non-Hermitian QuAPI~\cite{palmNonperturbativeEnvironmentalInfluence2017} is able to simulate the dynamics in presence of loss processes, it is unable to account for pumps or the impact of these losses on spectra. To alleviate these issues, one of us has
introduced a combination of path integrals and Lindblad master equation to
describe empirical effects.~\cite{boseIncorporationEmpiricalGain2024} The
path integral captures the impact of the dissipative environment on the quantum
system in a non-perturbative manner, and the Lindblad part deals with any and all processes that are
described at an empirical level. These empirical processes, including but not limited to processes
that pump or drain the system, are dealt with under a Markovian approximation. This path integral Lindblad
dynamics (PILD) method~\cite{boseIncorporationEmpiricalGain2024} has been used
to study linear spectra for lossy chiral aggregates both in and out of
polaritonic cavities as well.~\cite{sharmaImpactLossMechanisms2024}

Obtaining the time-evolving RDM is, however, only the first step towards
understanding any dynamics.
While, for a given initial location of the quantum particle, the RDM
can obviously give the time-dependent population on each site, it is often
extremely difficult to ascertain the route followed by the quantum particle
during the transport. A lot of effort has been focused on exploring these
aspects of the dynamics for ``closed'' systems without pumps or drains. Initial work using
flux networks and flux balance methods were used to analyze these pathways in
light-harvesting Fenna--Matthews--Olson complex
(FMO).~\cite{wuEfficientEnergyTransfer2012} More recently Dani and
Makri~\cite{daniCoherenceMapsFlowExcitation2023} have tried to understand the
dynamics using the concept of coherence maps. One of us has developed the related idea
of state-to-state transport~\cite{boseImpactSolventStatetoState2023, boseImpactSpatialInhomogeneity2023} which can
unravel the initial condition-dependent dynamical pathways that are present in
the system in a mathematically rigorous manner.
This state-to-state approach decomposes the dynamics into
direct unmediated transport flows between any two sites in presence of thermal
baths, thereby allowing us to identify the important connections in a dynamical manner,
and follow the quantum particle as it moves from one site to another over time.
These pairwise flows can then be pieced together to determine the routes of
transport.

Beyond these ``closed'' systems, to explore the mechanistic aspects of transport in systems with losses, we have
recently showed that one can extend the idea of state-to-state analysis to
simulations of non-Hermitian systems with
QuAPI.~\cite{sharmaNonHermitianState2025} Moreover, this extension allows
one to estimate transport efficiencies in aggregates with multiple points of
extraction in a sink-specific manner. However, we would now like to ask a
more general question --- is it possible to additionally incorporate pump processes in the state-to-state analysis as well? How do multiple sources and sinks interact with
the system in contact with the solvent environment? While non-Hermitian
path integrals cannot account for pumping processes, our path integral Lindblad dynamics can.

In this paper, we extend the state-to-state analysis to analyze the routes of
transport in systems subject to the simultaneous influence of solvent
environments, expressed as baths, and empirical processes encoded approximately
through Lindblad jump operators. As mentioned previously, such a division of the
different environments that the system is exposed to, allows us to treat the
solvent environments in a numerically exact manner while relegating the
empirical processes to approximate Lindblad master equation-based treatments.
Just as the vibrational and solvent degrees of freedom can change the details of
the pathways involved in the dynamics, so can the empirical processes that pump
or drain the system. Our present work is an attempt at unraveling these
mechanistic details. In a sense, the present Lindblad state-to-state method
describes a formalism that accounts for a superset of physical phenomena including pump processes in comparison to the previous
non-Hermitian idea.~\cite{sharmaNonHermitianState2025} Consistent with other
developments of the state-to-state method, our Lindblad state-to-state is
similarly independent of the exact method used to simulate the dynamics of the
system RDM so long as the impact of both the thermal bath and the empirical
processes is incorporated.


We start by developing the method in
Sec.~\ref{sec:method}, following which in Sec.~\ref{sub:nHcomp}, we demonstrate
its validity through a consistency check with the
previous non-Hermitian method for the case of an exciton-polariton system with
two competing loss mechanisms. We then apply the new method to explore systems
involving both pumps and drains (Sec.~\ref{sub:pump}
and~\ref{sub:pumpdrain}), and end with some concluding remarks in
Sec.~\ref{sec:conclusions}. A discussion on the limitations of the non-Hermitian description with respect to the Lindblad formalism is also presented in Appendix~\ref{app:nh_vs_lindblad}.

\section{Method}\label{sec:method}

Consider a $N$-state system coupled to a bath:
\begin{align}
    \hat{H} &= \hat{H}_\text{sys} + \hat{H}_\text{bath} + \hat{H}_\text{sys-bath}\label{eq:total_hamiltonian}
\end{align}
where, under Gaussian response theory, the solvent environment(s) has been mapped onto $N_b$ independent baths of harmonic oscillators~\cite{makriLinearResponseApproximation1999, boseZerocostCorrectionsInfluence2022},
\begin{align}
    \hat{H}_\text{bath} &= \sum_{s=1}^{N_b}\sum_b \frac{p_{sb}^2}{2}+\frac{1}{2}\omega_{sb}^2 x_{sb}^2 \\
    \hat{H}_\text{sys-bath} &= -\sum_{s=1}^{N_b}\sum_b c_{sb}x_{sb}\hat{S}_s
\end{align}
where the $j$th bath interacts with the system through the operator $\hat{S}_j$ and is characterized by the spectral density,
\begin{align}
    J_j(\omega) &= \frac{\pi}{2}\sum_b \frac{c_{jb}^2}{\omega_{jb}}\delta(\omega-\omega_{jb})~\label{eq:spect_dens}
\end{align}
obtained from molecular dynamics simulations or directly from experiments.
Often, for charge and exciton transport where the electronic states form the
system, nuclear motions of the molecules that are a part of the system and those that are a part of
the solvent both contribute to the bath degrees of freedom. Simulations of the
dynamics of this system-bath couple (Eq.~\ref{eq:total_hamiltonian}) scales
exponentially with the dimensionality of the full Hilbert space, and
consequently becomes intractable. Therefore, simulations are done for the reduced density
matrix corresponding to the system, $\rho_\text{sys}(t)$.

Methods based on Feynman-Vernon influence functional path integrals provide a numerically exact approach for simulating these systems interacting with potentially multiple baths~\cite{feynmanTheoryGeneralQuantum1963}. For a separable initial condition $\rho(0)=\rho_\text{sys}(0)\otimes e^{-\beta\hat{H}_\text{bath}}/Z_\text{bath}$, the time-evolved reduced density matrix corresponding to the system can be written as~\cite{makriTensorPropagatorIterativeI1995, makriTensorPropagatorIterativeII1995}:
\begin{widetext}
\begin{align}
    \mel{s_N^+}{\rho_\text{sys}(N\Delta t)}{s_N^-} &= \sum_{s_0^\pm}\sum_{s_1^\pm}\ldots\sum_{s_{N-1}^\pm} \mel{s_N^\pm}{\mathcal{E}_0(\Delta t)}{s_{N-1}^\pm}\mel{s_{N-1}^\pm}{\mathcal{E}_0(\Delta t)}{s_{N-2}^\pm}\ldots\mel{s_{1}^\pm}{\mathcal{E}_0(\Delta t)}{s_{0}^\pm} \nonumber \\
    &\times\mel{s_0^+}{\rho_\text{sys}(0)}{s_0^-} F\left[\{s_j^\pm\}\right]\label{eq:fvpi}
\end{align}
\end{widetext}
where $\mathcal{E}_0(\Delta t)=e^{-i\hat{H}_\text{sys}\Delta t/\hbar}\otimes e^{i\hat{H}_\text{sys}\Delta t/\hbar}$ is the dynamical map corresponding to the bare system, $s_j^\pm$ is the state of the system at the $j$th time point and $F\left[\{s_j^\pm\}\right]$ is the Feynman-Vernon influence functional along the path $\{s_j^\pm\}$~\cite{feynmanTheoryGeneralQuantum1963}. The influence functional $F$ is dependent on the bath response function and consequently also on the spectral density as specified in Eq.~\ref{eq:spect_dens}. It accounts for the non-Markovian effects of the environment and can be calculated analytically for harmonic baths~\cite{makriTensorPropagatorIterativeI1995} or estimated using semiclassical or classical trajectories~\cite{makriSemiclassicalInfluenceFunctional1998, lambertQuantumclassicalPathIntegralI2012, lambertQuantumclassicalPathIntegralII2012} for atomistic baths.

Additionally, to complete the description of the problem, often the system-bath couple (Eq.~\ref{eq:total_hamiltonian}) is not isolated. It interacts with a larger universe and is, consequently, open to other processes (besides the rigorously described interactions with the bath). They may cause changes in the state of the system like spontaneous emission, pumping/loss of quantum particles, etc. While one can attempt to describe the atomistic details of such processes, this becomes an exceedingly challenging task. As long as one is interested solely in their impact on the preceding transport, such a careful parameterization of these processes may be unnecessary. One can choose to include these processes on an ``empirical'' level by incorporating a rough time-scale through Lindblad jump operators that interact with the system. Without loss of generality, one can presume that the Lindblad jump operators are
constructed as a sum of elementary jump operators of the form:
\begin{align}
    L_n &= T_n^{-\frac{1}{2}}\sum_j \tilde L_{nj}\\
    \tilde L_{nj} &= c_{nj}\dyad{f_{nj}}{i_{nj}}
\end{align}
where $T_n$ is the time-scale of action of the $n$th jump operator, $c_{nj}$ is
some coefficient and $\ket{i_{nj}}$ and $\ket{f_{nj}}$ are system states. (On a cautionary note, while this might be a fair zeroth order approximation to treat the ``external interactions'' through Lindblad jump operators, because we are primarily interested in the transport in the aggregate, one cannot, in general, treat the influence of vibrational and solvent degrees of freedom on the system using Lindblad operators. The Lindblad quantum master equation is an intrinsically Markovian equation of motion and is consequently unable to capture the non-Markovian and non-perturbative influence of the bath as shown in Eq.~\ref{eq:fvpi}.)

In the presence of such external empirical processes, the equation of motion for the reduced density matrix of such a system-bath set is given by the Lindblad master equation~\cite{lindbladGeneratorsQuantumDynamical1976, goriniCompletelyPositiveDynamical1976}:
\begin{align}
    \dot\rho(t) &= -\frac{i}{\hbar}\comm{\hat{H}}{\rho(t)} \nonumber\\
    &+ \sum_n \left(L_n\rho(t)L_n^\dag - \frac{1}{2}\acomm{L_n^\dag L_n}{\rho(t)}\right).\label{eq:lindbladME}
\end{align}
where $\rho(t)$ is a function of the system ($s^\pm$) and bath ($x^\pm$) degrees of freedom, that is, $\rho(t) = \sum_{s^\pm}\idotsint\dd{x^\pm} \ket{s^+, x^+}\mel{s^+, x^+}{\rho(t)}{s^-, x^-}\bra{s^-, x^-}$.

At this stage, it is useful to emphasize the triple-layered nature of the full
universe that is involved in this description --- we have the system layer, the
bath comprising of the vibrational and solvent degrees of freedom, and the
external universe which interacts with the system through the Lindblad jump
operators. Na\"ively solving Eq.~\ref{eq:lindbladME} while propagating both the system and bath degrees of freedom in an exact manner is computationally infeasible because of the exponential scaling with respect to the dimension of the system-bath Hilbert space. Our recently developed PILD
method~\cite{boseIncorporationEmpiricalGain2024} offers a convenient approach to extending QuAPI~\cite{makriTensorPropagatorIterativeI1995, makriTensorPropagatorIterativeII1995} to incorporate empirical Lindblad operators in addition to the thermal baths. It, therefore, allows for the simulation of the RDM corresponding to the system,
$\rho_\text{sys}(t) = \Tr_\text{bath}\left[\rho(t)\right]$, where $\rho(t)$ is
the RDM corresponding to the system-bath portion satisfying Eq.~\ref{eq:lindbladME} with $\rho(0)=\rho_\text{sys}(0)\otimes e^{-\beta\hat{H}_\text{bath}}/Z_\text{bath}$.

According to PILD,~\cite{boseIncorporationEmpiricalGain2024}
\begin{align}
    \dot\rho_\text{sys}(t)&=\mathcal{E}_0^{-1}(t)\left(\vphantom{\sum_n}\int_0^{\tau_\text{mem}}\mathcal{K}(\tau)\rho_\text{sys}(t-\tau)d\tau\right. \nonumber \\
    &+\left. \sum_n \left(L_n\rho_\text{sys}(t)L_n^\dag - \frac{1}{2}\acomm{L_n^\dag L_n}{\rho_\text{sys}(t)}\right)\right)\label{eq:pild}
\end{align}
where $\mathcal{K}(\tau)$ is the non-Markovian memory kernel and $\tau_\text{mem}$ is the memory length. This memory kernel can be obtained accurately from approximate~\cite{mulvihillCombiningMappingHamiltonian2019,mulvihillModifiedApproachSimulating2019} or numerically exact~\cite{chatterjeeRealTimePathIntegral2019} simulations of the time evolution of $\rho_\text{sys}(t)$ in the absence of the Lindbladians. Alternatively, one can use the transfer tensor method~\cite{cerrilloNonMarkovianDynamicalMaps2014} to link the dynamical map of the reduced system in presence of the solvent environment, $\mathcal{E}(t)$ where $\rho_\text{sys}(t) = \mathcal{E}(t)\rho_\text{sys}(0)$, to transfer tensors, that are analogous to time-discretized versions of the memory kernel. One can now use any path integral method to simulate the dynamical map of the system-bath set in absence of the Lindbladians and extract from there the solvent memory kernel. (It is trivial to modify Eq.~\ref{eq:fvpi} to yield the dynamical maps by removing the sum over $s^\pm_0$ and removing the initial reduced density matrix factor.) On solving Eq.~\ref{eq:pild} with the desired Lindblad jump operators, one obtains the combined effect of the non-Markovian solvent environment and the Markovian empirical Lindblad processes.

In this paper we ask a slightly different question: beyond the population dynamics obtained from $\rho_\text{sys}(t)$, is it possible to inquire into the routes of transport that the system shows in presence of both the bath and the external empirical process? Our state-to-state analysis framework~\cite{boseImpactSolventStatetoState2023} provides a way to answer this question. Here we try to apply the same logic to the system-bath problem subject to empirical pumps and drains.

We start by analyzing the rate of change of the population of a particular system state. Using Eq.~\ref{eq:lindbladME}, the time derivative of the population of system state $\ket{l}$ can be written as:
\begin{align}
    \dot P_l(t) &= \mel{l}{\dot{\rho}_\text{sys}(t)}{l} \nonumber \\
    &= \mel{l}{\Tr_\text{bath}\left[\dot\rho(t)\right]}{l} \nonumber \\
    &= \dot{P}_l^H(t) + \dot{P}_l^L(t) \label{eq:pldot}
\end{align}
where we have split the expression between terms arising out of the commutator with the Hamiltonian, $\dot{P}_l^H(t)$, and ones arising from the Lindbladian terms, $\dot{P}_l^L(t)$. Therefore,
\begin{widetext}
\begin{align}
    \dot{P}^H_l(t) &= -\frac{i}{\hbar}\mel{l}{\Tr_\text{bath}\left(\comm{\hat{H}}{\rho(t)}\right)}{l} \nonumber \\
    &= -\frac{i}{\hbar}\sum_r\left(\mel{l}{\hat{H}_\text{sys}}{r}\mel{r}{\rho_\text{sys}(t)}{l}-\mel{l}{\rho_\text{sys}(t)}{r}\mel{r}{\hat{H}_\text{sys}}{l}\right) \label{eq:H_part} \\
    \dot{P}^L_l(t) &= \sum_n \mel{l}{\left(L_n\rho_\text{sys}(t)L_n^\dag - \frac{1}{2}\acomm{L_n^\dag L_n}{\rho_\text{sys}(t)}\right)}{l} \nonumber \\
    &= \sum_n T_n^{-1}\sum_{j,k}c_{nj}c_{nk}^{*}\left( \vphantom{\frac{1}{2}} \mel{i_{nj}}{\rho_\text{sys}(t)}{i_{nk}}\delta_{l,f_{nj}}\delta_{l,f_{nk}} - \frac{1}{2}\delta_{f_{nj}, f_{nk}}\left(\delta_{l,i_{nj}}\mel{i_{nk}}{\rho_\text{sys}(t)}{l} + \delta_{l,i_{nk}}\mel{l}{\rho_\text{sys}(t)}{i_{nj}}\right)\right)\label{eq:L_part}
\end{align}
\end{widetext}
The simplification done in Eq.~\ref{eq:H_part} is only possible in the case of diagonal system-bath coupling operator, $\hat{H}_\text{sys-bath}$. This is consistent
with the Frenkel-Holstein model of exciton dynamics or the system-bath
decompositions conventionally used to describe charge transfer
processes. Equation~\ref{eq:L_part} uses the fact that the Lindblad jump operators considered act only
on the Hilbert space of the system. Thus, in representing both the Hamiltonian
and the Lindbladian contributions to the rate of change in terms of
$\rho_\text{sys}(t)$, we have been able to relax our requirement of knowing the
dynamics corresponding to the system-bath couple. Now, $\dot{P}_l(t)$ in
Eq.~\ref{eq:pldot} can be computed from $\rho_\text{sys}(t)$ and
$\hat{H}_\text{sys}(t)$. Therefore, methods like PILD that simulate
$\rho_\text{sys}(t)$ directly can be used to obtain the relevant dynamics.

Now to complete the state-to-state analysis, we need to be able to partition the rate of change of population of state $\ket{l}$ in terms of other states $\ket{r}$:
\begin{align}
    \dot{P}_l(t) &= \sum_r \dot{P}_{l\leftarrow r}(t).\label{eq:basestate2state}
\end{align}
where $\dot{P}_{l\leftarrow r}(t)$ is the instantaneous rate of change of the population of the $l$th site due to the $r$th site. This is achieved trivially for $\dot{P}_l^H(t)$, which already has this structure in Eq.~\ref{eq:H_part}. However, notice that for $\dot{P}_l^L(t)$ in Eq.~\ref{eq:L_part}, this partitioning is not possible. There are summations over two states, $\ket{i_{nk}}$ and $\ket{i_{nj}}$ for each  Lindbladian $L_n$ which may contribute to the flux of system state $\ket{l}$. How does one choose a unique system state $\ket{r}$ and decompose the population flux as in Eq.~\ref{eq:basestate2state}?

The ambiguity and complexity stemming from the multiple states $\ket{i_{nk}}$, $\ket{i_{nj}}$, and the extra summation over different Lindbladians can all be traced to our definitions of the jump operators as an
unconstrained sum of elementary Lindbladians. Such a description
is probably more general than required for dealing with simple empirical pumping
and draining processes. To understand the action of multiple such processes and the practical
restrictions that one might impose on them, consider an exciton transport system
with multiple extraction points. The key idea we want to consider is the spatial locality of the empirical processes. If we drain or pump the $j$th site through a process, a different site $k$ would typically not be affected. This means that the initial state that a particular process acts on uniquely defines the final state. Suppose the process under consideration is draining a site. In this case, it uniquely takes the first excited state of this molecule to the ground state. On the other hand, if this was a pumping process, it will uniquely take this molecule from the ground to the excited state. Of course, the full many-body operator would be defined as the direct product of this one-body operator with identities on all the other sites. This is what we want to encode in our empirical Lindbladians. 

Motivated by the site- and state- specific nature of pumping and draining processes, we put a single, probably weaker, physical
restriction on them: if two elementary jump operators $\tilde L_{nj}$ and $\tilde L_{nj'}$ are part of a single
jump operator $L_n$ (signifying the $n$th process), then they cannot have the same end-point. Mathematically, if
$j\ne k$, then for all $n$, $\ket{f_{nj}}\ne\ket{f_{nk}}$. In other words, a
single process $L_n$ should not map two different initial states to the same final
state. Processes mapping different initial states (whether on the same or different sites) to the same final state would, in a plurality of cases, be better represented by two different Lindbladians $L_n$ and $L_{n'}$ owing to different external environment modes accounting for them. This serves as the minimal criterion that can account for most such pumping/draining processes, and also gets rid of the double summation over $j$ and $k$ in Eq.~\ref{eq:L_part}. Thus, upon simplification, Eq.~\ref{eq:pldot} gives:
\begin{align}
    \dot{P}_l(t) &= -\frac{2}{\hbar}\sum_r\mel{l}{\hat{H}_\text{sys}}{r}\Im\mel{l}{\rho_\text{sys}(t)}{r}\nonumber\\
    &+ \sum_n T_n^{-1}\sum_j\abs{c_{nj}}^2\mel{i_{nj}}{\rho_\text{sys}(t)}{i_{nj}}\left(\delta_{l,f_{nj}} - \delta_{l,i_{nj}}\right).\label{eq:derivative_rdm_simple}
\end{align}
In the light of Eq.~\ref{eq:derivative_rdm_simple}, let us now rewrite the Lindbladian contribution to the population flux:
\begin{align}
    \dot{P}^L_l(t) &= \sum_n T_n^{-1}\sum_j\abs{c_{nj}}^2\mel{i_{nj}}{\rho_\text{sys}(t)}{i_{nj}}\left(\delta_{l,f_{nj}} - \delta_{l,i_{nj}}\right) \label{eq:final_lindbladian_part}
\end{align}
To understand how to decompose it in a state-specific manner, consider the effect of a single elementary Lindbladian $T^{-1/2}\dyad{f}{i}$, which causes
population to flow from the $i$th state to the $f$th state. Below are the rates
of change of these two states caused only by the Lindbladian terms. (The
Hamiltonian part has been suppressed.)
\begin{align}
    \dot{P}_f^L(t) &= T^{-1}\mel{i}{\rho_\text{sys}(t)}{i}\label{eq:fthstate}\\
    \dot{P}_i^L(t) &= -T^{-1}\mel{i}{\rho_\text{sys}(t)}{i}\label{eq:ithstate}.
\end{align}
The origins of Eqs.~\ref{eq:fthstate} and~\ref{eq:ithstate}
are in the first and second Lindbladian terms of
Eq.~\ref{eq:final_lindbladian_part} respectively. As expected, the
rate of change of the population of the $i$th state is negative, and that of the
$f$th state is positive reflecting the direction of the population flow. Now to
assign the ``source'' of these changes, notice that the Lindbladian causes
population to flow from $\ket{i}$ to $\ket{f}$. Consequently, $\dot{P}_i^L(t)$
must be caused by the $f$th state, and $\dot{P}_f^L(t)$ must have as its source
the $i$th site. Thus, the source-resolved population flux equation would be given by:
\begin{align}
    \dot{P}_{l\leftarrow r}(t) &= -\frac{2}{\hbar}\mel{l}{\hat{H}_\text{sys}}{r}\Im\mel{l}{\rho_\text{sys}(t)}{r}\nonumber\\
    + \sum_n T_n^{-1}&\sum_j\abs{c_{nj}}^2\mel{i_{nj}}{\rho_\text{sys}(t)}{i_{nj}}\left(\delta_{l,f_{nj}}\delta_{r, i_{nj}} - \delta_{l,i_{nj}}\delta_{r, f_{nj}}\right)\label{eq:derivative_rdm_decomposed}
\end{align}

The final step is where the expressions are integrated to obtain the direct and
unmediated transport from a state $\ket{r}$ to a state $\ket{l}$.
\begin{align}
    P_{l\leftarrow r}(t) &= -\frac{2}{\hbar}\mel{l}{\hat{H}_\text{sys}}{r}\int_0^t\dd{t'}\Im\mel{l}{\rho_\text{sys}(t')}{r}\nonumber\\
    &+ \sum_n T_n^{-1}\sum_j\int_0^t\dd{t'}\mel{i_{nj}}{\rho_\text{sys}(t')}{i_{nj}}\abs{c_{nj}}^2\delta_{l,f_{nj}}\delta_{r,i_{nj}}\nonumber\\
    &- \sum_n T_n^{-1}\sum_j\int_0^t\dd{t'}\mel{i_{nj}}{\rho_\text{sys}(t')}{i_{nj}}\abs{c_{nj}}^2\delta_{l,i_{nj}}\delta_{r,f_{nj}}\label{eq:state2state}
\end{align}
This is the final form of the Lindblad state-to-state formalism. The first term of Eq.~\ref{eq:state2state}
is exactly the same as the traditional state-to-state
transport~\cite{boseImpactSolventStatetoState2023}. It accounts for the rate at
which the population is transferred from $\ket{r}$ to $\ket{l}$ via the
Hamiltonian. This we will call the Hamiltonian transport or the Hamiltonian
flow. Additional transport happens through the jump operators, which we will
call the Lindbladian transport or flow.

Next consider the two Lindbladian transport terms in  Eq.~\ref{eq:state2state} separately. They show
the total change in the population of the $l$th state due to various jump
operators, if the ending state is $\ket{l}$ and the starting state is $\ket{r}$.
Specifically, the first Lindbladian term talks about the total increase in the population
of the $l$th site because of the $r$th site through all the Lindblad mechanisms
while the second term talks about the total decrease in the population of the
$l$th site because of the total Lindbladian flow from the $l$th to the $r$th
site. These two terms together constitute the net Lindbladian transport.

Equation~\ref{eq:state2state} can be shown to be consistent with the principle
of detailed balance, $P_{l\leftarrow r}(t) = - P_{r\leftarrow l}(t)$. Also,
$P_{l\leftarrow l}(t) = 0$ showing that no self-transfer can happen. Moreover,
since Eq.~\ref{eq:state2state} is expressed in terms of $\hat{H}_\text{sys}$ and
$\rho_\text{sys}(t)$, it remains valid even for the case when no explicit bath is there. For the purposes of this paper, we are
going to concentrate on pumping and draining processes acting on a system-bath
couple, and numerically explore the consequences of Eq.~\ref{eq:state2state}.

\section{Numerical Results}\label{sec:result}

We will demonstrate the Lindblad state-to-state analysis method through a series of examples pertaining to
polaritonic and excitonic dynamics. First, in Subsection~\ref{sub:nHcomp}, we
will demonstrate the consistency of the current method with the recently developed
non-Hermitian state-to-state~\cite{sharmaNonHermitianState2025} analysis method
for the set of mutually applicable problems. Then we move on to examples where
the non-Hermitian state-to-state analysis is not applicable --- we demonstrate
the Lindblad state-to-state method using the case of an excitonic dimer being pumped
(Subsection~\ref{sub:pump}), and then simultaneously pumped and drained from
different sites (Subsection~\ref{sub:pumpdrain}). We use our PILD
method~\cite{boseIncorporationEmpiricalGain2024, sharmaImpactLossMechanisms2024}
implemented through the \texttt{QuantumDynamics.jl}
package~\cite{boseQuantumDynamicsJlModular2023} to obtain $\rho_\text{sys}(t)$
for all simulations presented herein. The time-evolved matrix product operators
(TEMPO)~\cite{strathearnEfficientNonMarkovianQuantum2018} implementation of
QuAPI is used to simulate the dynamical maps required for PILD.

\subsection{Comparison with non-Hermitian State-to-State}\label{sub:nHcomp}
Both the current Lindblad state-to-state approach and the recently published
non-Hermitian state-to-state approach~\cite{sharmaNonHermitianState2025} seem to
enable exploration of routes of transport and transport efficiencies for open
quantum systems with empirically described loss processes. In the previous work,
those processes were described by non-Hermiticities, whereas here, they are
described by the Lindblad jump operators. While the non-Hermitian method may be
able to get away with a smaller system dimensionality in certain cases, the
Lindblad approach is significantly more general as we shall demonstrate through
later numerical examples. (This difference of the Lindblad approach over the non-Hermitian description is also discussed in Appendix~\ref{app:nh_vs_lindblad}.) The common pool of problems that both the approaches can deal with are cases
where the only empirical processes impacting the system are one or more loss (drain)
sites. We, therefore, use such a
case to validate our Lindblad state-to-state method.

It should be noted at the outset that both the methods have different empirical ways of incorporating the losses. There cannot be a guarantee of getting identical results. The check, therefore, is one of consistency of conclusions obtained from either methods. In the limit of extremely weak empirical processes both methods should become identical. In our numerical exploration, we show that surprisingly all the observables turn out to be the same between the non-Hermitian and the Lindbladian treatments.

Consider a nearest-neighbor polaritonic trimer where an excitonic trimer is
coupled to a Fabry-P\'erot cavity mode:
\begin{align}
    \hat{H}_\text{sys} &= \epsilon_0\dyad{0} + \sum_j\epsilon_j\dyad{j} + \sum_{j<k} h_{jk}\left(\dyad{j}{k} + \dyad{k}{j}\right) \nonumber \\ &\hspace{1em}+\hbar\omega_c\dyad{c} + \sum_j\Omega\left(\dyad{j}{c}+\dyad{c}{j}\right),
\end{align}
where $\ket{0}$ is the ground state of the system, $\ket{j}$ represents the state where the excitation is on the $j$th monomer and $\ket{c}$ is the cavity mode. The parameters are taken to be same as one of the examples from Ref.~\citenum{sharmaNonHermitianState2025}, which we summarize here for convenience. The energy of the ground state $\epsilon_0$ is taken to be $\SI{0}{\per\cm}$. All the monomers are assumed to be identical with the same excitation energies. Therefore, $\epsilon_j$ is independent of the site number $j$. Because, we are interested in the dynamics starting from a first-excitation subspace state, we can set $\epsilon_j=0$ for all $j$. The nearest-neighbor inter-monomer coupling is taken as $h_{j,k}=-h=\SI{-181.5}{\per\cm}\delta_{k,j+1}$. The exciton is harvested from the third monomer or $\ket{3}$ with a time-scale of $T_3=\SI{300}{\fs}$. The cavity is taken to be resonant with the molecular Frank-Condon excitation energy. All Fabry-P\'erot cavities, typically, have a certain time-scale with which they lose the photon. Such a loss diverts a part of the excitonic transport, as the photon is never truly harvested. Following the example in Ref.~\citenum{sharmaNonHermitianState2025}, we take the time-scale of loss from cavity to be $T_c=\SI{600}{\fs}$, and model the molecular nuclear environment for each monomer using the Ohmic spectral density:
\begin{align}
   J(\omega) = 2\pi\hbar\,\xi\omega\exp\left(-\omega/\omega_\text{cutoff}\right),\label{eq:spectraldensity}
\end{align}
where the Kondo parameter $\xi=0.121$ and
$\omega_\text{cutoff}=\SI{900}{\per\cm}$ corresponding to a reorganization
energy $\lambda_0=\SI{217.8}{\per\cm}$. The cavity mode is of course not
associated with any bath. All the simulations are done at a temperature of
$\SI{300}{\kelvin}$.

\begin{figure}
    \centering
    \includegraphics{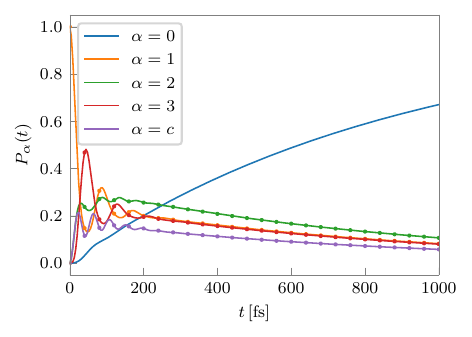}
    \caption{Population, $P_\alpha(t)$, of different states $\ket{\alpha}$ in a polaritonic trimer with an initial excitation $\rho_\text{sys}(0)=\dyad{1}$. (Markers: non-Hermitian state-to-state results~\cite{sharmaNonHermitianState2025}; Lines: Lindblad state-to-state results)}
    \label{fig:pop_trimer_cav}
\end{figure}

The losses on the third monomer and the cavity are incorporated in different ways for the two methods. For the non-Hermitian system, $\epsilon_3$ and $\omega_c$ are made complex with $\Im(\epsilon_3)=-\pi\hbar/T_3$ and $\Im(\omega_c)=-\pi/T_c$.~\cite{sharmaNonHermitianState2025} The Lindbladian description is more elaborate. The two losses are accounted for by two different jump operators
\begin{align}
    L_3 &= T^{-1/2}_3 \dyad{0}{3},\\
    L_c &= T^{-1/2}_c \dyad{0}{c}.
\end{align}
Notice that unlike the non-Hermitian case where the target site of the losses is undetermined, for the Lindblad method we explicitly state that both the jump operators bring the system down into the same state $\ket{0}$. (Note that $\ket{0}$ is not even included in the non-Hermitian calculations.) We first show the dynamics obtained from both the methods in Fig.~\ref{fig:pop_trimer_cav}, which match exactly. A time-step of $\Delta t=\SI{4}{\fs}$ and a non-Markovian memory length of $\tau_\text{mem}=\SI{200}{\fs}$ (amounting to 50 time-steps) were used for the converged dynamics in both cases. However, the increase of the ground state population due to the losses is a feature that is captured only by the PILD method~\cite{boseIncorporationEmpiricalGain2024}.

\begin{figure}
    \centering
    \hspace*{-0.5cm}
    \subfloat[Flow into monomer 1]{\includegraphics{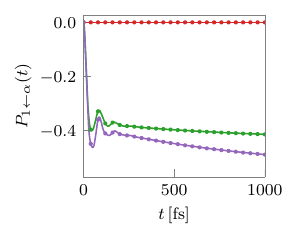}}
    \subfloat[Flow into monomer 2]{\includegraphics{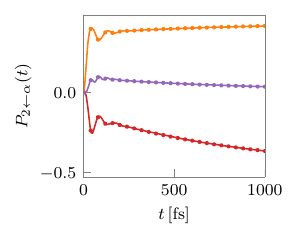}}

    \hspace*{-0.5cm}
    \subfloat[Flow into monomer 3]{\includegraphics{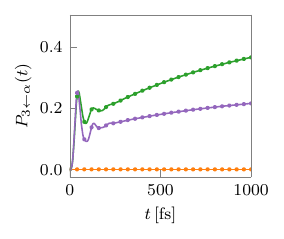}}
    \subfloat[Flow into cavity]{\includegraphics{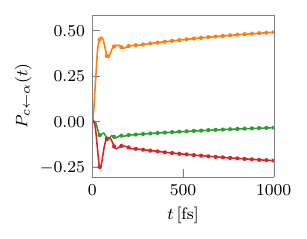}}
    
    \caption{State-to-state analysis of excitation flows into different sites of the lossy polaritonic trimer. (Lines: Lindblad state-to-state result; Markers: non-Hermitian state-to-state results~\cite{sharmaNonHermitianState2025}; Orange: $P_{*\leftarrow 1}(t)$, Green: $P_{*\leftarrow 2}(t)$, Red: $P_{*\leftarrow 3}(t)$ and Purple: $P_{*\leftarrow c}(t)$. Terms of the type $P_{\alpha\leftarrow\alpha}(t)$ are not depicted here.)}
    \label{fig:state}
\end{figure}

\begin{figure}
    \centering
    \includegraphics{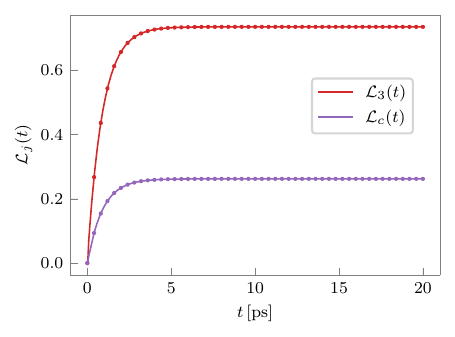}
    \caption{Site-specific excitation loss, $\mathcal{L}_{j}(t)$, from the polaritonic trimer. Lines: Lindblad state-to-state, Markers: non-Hermitian method from Ref.~\citenum{sharmaNonHermitianState2025}.}
    \label{fig:loss}
\end{figure}

Next, we compare the state-to-state analysis obtained from both the methods in Fig.~\ref{fig:state}, excluding the loss terms. We have previously
analysed the physics of this problem in
depth~\cite{sharmaNonHermitianState2025} and avoid going into the details.
Notice that the results obtained from the Lindblad state-to-state method matches
those obtained from our non-Hermitian state-to-state method.

There is, however, an intrinsic difference in the interpretation of the loss
terms between the non-Hermitian state-to-state approach and the current
approach. As discussed, the non-Hermitian approach does not necessitate the
inclusion of the ground state $\ket{0}$; loss, $\mathcal{L}_j(t)$, from a site
$j$ is given by $\abs{P_{j\leftarrow j}(t)}$. However, for the Lindblad
state-to-state picture (Eq.~\ref{eq:state2state}), it can be trivially shown
that $P_{j\leftarrow j}(t)=0$. Loss, here, is seen as a Lindbladian transport of the
system from the $j$th site to the ground state, $\mathcal{L}_j(t) =
P_{0\leftarrow j}(t)$. In Fig.~\ref{fig:loss}, we show the loss from
$\ket{j}$ for $j=3,c$ using both the methods, which are once again identical.
Additionally, as expected, there is no loss from sites 1 and 2 into the ground
state (not included therefore in Fig.~\ref{fig:loss}).

This example demonstrates the equivalence of the Lindblad state-to-state method and
the non-Hermitian state-to-state method for the subset of problems where both
methods are applicable.

\subsection{Pumped Excitonic Dimer}\label{sub:pump}

Next we move onto the first case where the current method is uniquely
applicable. Imagine an excitonic dimer that is initially in the ground state.
The left monomer (monomer 1) is pumped incoherently with a particular time-scale,
$T_\mathrm{pump}$. We want to understand the flow of excitation into this system. Unlike in
our previous example, this system can no longer be described using the first
excitation subspace. That is because the number of excitation keeps rising till
we have two excitations (or $N$ excitations for an $N$-mer in general). 

The full space Hamiltonian for an $N$-mer with identical monomers and only nearest-neighbor couplings, $h_{j,k}=-h=\SI{-181.5}{\per\cm}\delta_{k,j+1}$, can be written in terms of the localized diabatic basis formed by
the direct products of $\ket{g_j}$ and $\ket{e_j}$ denoting the molecular ground and
excited states, respectively, on the $j$th monomer as:
\begin{align}
    \hat{H}_\text{sys} &= \epsilon\sum_{j=1}^N\dyad{e_j}{e_j} \nonumber\\
    &\hspace{1em}-h\sum_{j=1}^{N-1}\left(\dyad{e_j g_{j+1}}{g_j e_{j+1}}+\dyad{g_j e_{j+1}}{e_j g_{j+1}}\right)\label{eq:exciton_fullspace}
\end{align}
where $\epsilon=\SI{1000}{\per\cm}$ is the monomeric excitation energy. Each of
these monomers are once again coupled to the same Ohmic vibrational bath as
before (Eq.~\ref{eq:spectraldensity}). For such an $N$-mer, any pump on a site $j$ can be written as $L_j^\text{pump}=T_\text{pump}^{-1/2}\dyad{e_j}{g_j}$ while a drain on site $j$ can be written as $L_j^\text{drain}=T_\text{drain}^{-1/2}\dyad{g_j}{e_j}$. 

For the excitonic dimer defined using Eq.~\ref{eq:exciton_fullspace} for $N=2$, the excitation pump at the left monomer can be written in an expanded fashion as $L_1^\text{pump}=T_\text{pump}^{-1/2}\left(\dyad{eg}{gg} +
\dyad{ee}{ge}\right)$, where $T_\text{pump}=\SI{300}{\fs}$. The two terms in the Lindbladian
correspond to a pumping into the first excitation subspace from the ground
state, and from the first excitation subspace to the doubly excited state
respectively.

\begin{figure}
    \centering
    \includegraphics{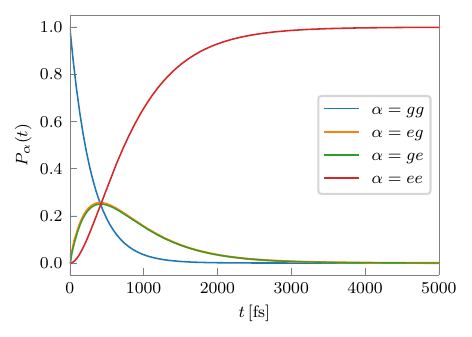}
    \caption{Population, $P_\alpha(t)$, of different states $\ket{\alpha}$ in the excitonic dimer initially in the ground state, $\rho_\text{sys}(0)=\dyad{gg}$, with excitation being pumped into monomer 1 with a time-constant of $T_\mathrm{pump}=\SI{300}{\fs}$.}
    \label{fig:pump}
\end{figure}

\begin{figure}
    \centering
    \hspace*{-0.5cm}
    \subfloat[Flow into $\ket{gg}$]{\includegraphics{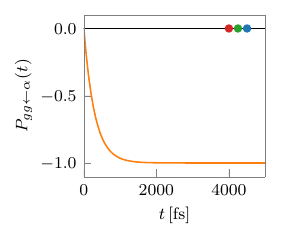}}
    \subfloat[Flow into $\ket{eg}$]{\includegraphics{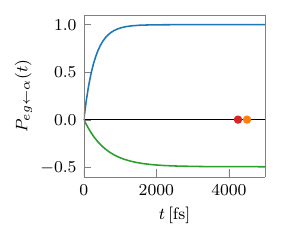}}

    \hspace*{-0.5cm}
    \subfloat[Flow into $\ket{ge}$]{\includegraphics{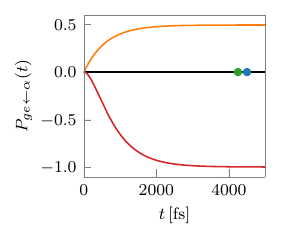}}
    \subfloat[Flow into $\ket{ee}$]{\includegraphics{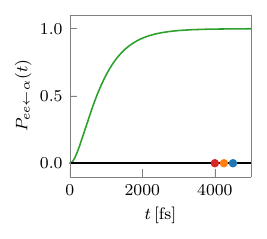}}
    
    \caption{State-to-state analysis of excitation flows into the diabatic states of the excitonic dimer when pumped with $T_\mathrm{pump}=\SI{300}{\fs}$. (Blue: $P_{*\leftarrow gg}(t)$, Orange: $P_{*\leftarrow eg}(t)$, Green: $P_{*\leftarrow ge}(t)$ and Red: $P_{*\leftarrow ee}(t)$. Black indicates multiple overlapping curves with legends marked as discs of corresponding colors.)}
    \label{fig:state_pump}
\end{figure}

The dynamics from an initially unexcited excitonic dimer, $\rho_\text{sys}(0) =
\dyad{gg}$, is shown in Fig.~\ref{fig:pump}. Convergence was
reached at a time-step of $\Delta t=\SI{4}{\fs}$ and a memory time of
$\tau_\text{mem}=\SI{400}{\fs}$. The system starts from $\ket{gg}$, but soon
gains excitation. Because there are no excitonic drains, eventually all the
population moves into the doubly excited state $\ket{ee}$. The population of
$\ket{eg}$ increases slightly before $\ket{ge}$ because it is the state that is
getting pumped. The state $\ket{ge}$ gains population through a Hamiltonian
transfer from the $\ket{eg}$ state because of the nearest-neighbor coupling.

The state-to-state analysis in the diabatic basis for this pumped dimer is shown
in Fig.~\ref{fig:state_pump}. Consistent with the Lindblad jump operators,
the pumping procedure takes the system from $\ket{gg}$ to 
$\ket{eg}$. The $\ket{ge}$ state only receives population from the
$\ket{eg}$ state. There is also only a single route of transport into 
$\ket{ee}$, which is from $\ket{ge}$ and Lindbladian in origin.

\begin{figure}
    \centering
    \includegraphics{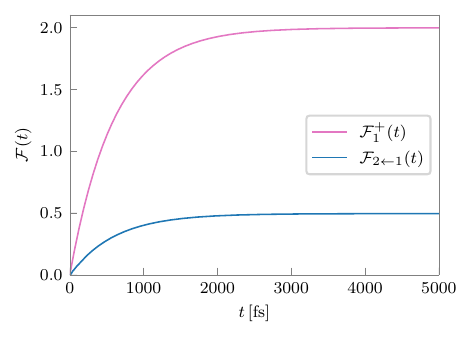}
    \caption{Total excitation flows $\mathcal{F}(t)$ into ($\mathcal{F}^{+}_{\alpha}(t)$) and between ($\mathcal{F}_{\alpha\leftarrow\beta}(t)$) monomers $\alpha$ (and $\beta$) of the excitonic dimer being pumped from monomer $1$ with corresponding timescale $T_\mathrm{pump}=\SI{300}{\fs}$.}
    \label{fig:site_flows}
\end{figure}


    

Finally, in these kinds of problems, it is also interesting to think about the
excitation flows not between the diabatic states, but in terms of the molecules.
We would, for instance, like to ask what is the net flow of exciton into the
first monomer. The diabatic state-to-state analysis discussed above gives a
perfect starting point for answering these questions. The flow of excitation
into the first monomer, $\mathcal{F}^+_1(t)$, through the pumping mechanism can
be trivially shown to be $P_{eg\leftarrow gg}(t) + P_{ee\leftarrow ge}(t)$.  In
a similar manner, the flow of excitation from the first to the second monomer,
$\mathcal{F}_{2\leftarrow1}(t)$, is just $P_{ge\leftarrow eg}(t)$. These flows
are shown in Fig.~\ref{fig:site_flows}. The flow into the first monomer is
positive as is the flow from monomer 1 to 2.

\subsection{Simultaneous Pumping and Draining}\label{sub:pumpdrain}

\begin{figure}
    \centering
    \includegraphics{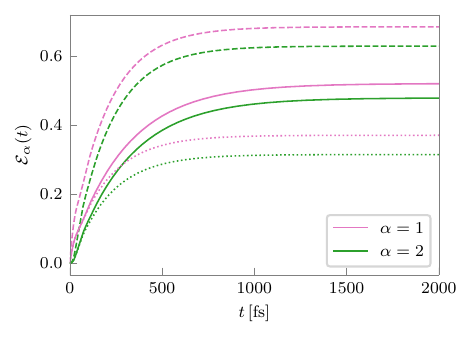}
    \caption{Total excitation $\mathcal{E}_{\alpha}(t)$ accumulated on the monomer $\alpha$ in the excitonic dimer being pumped and drained simultaneously from monomers $1$ and $2$, respectively, with corresponding timescales $(T_\text{pump},T_\text{drain})$ of $(\SI{150}{\fs},\SI{300}{\fs})$ (dashed), $(\SI{300}{\fs},\SI{300}{\fs})$ (solid) and $(\SI{300}{\fs},\SI{150}{\fs})$ (dotted).}
    \label{fig:excitation_dimer}
\end{figure}

\begin{figure}
    \centering
    \includegraphics{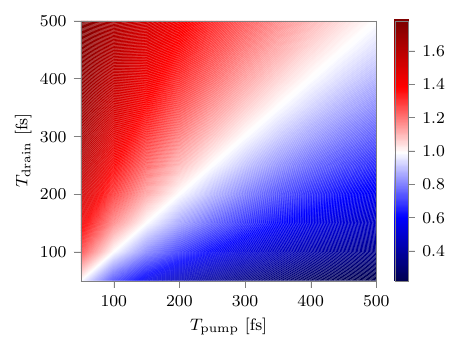}
    \caption{Total steady-state excitation accumulated in the excitonic dimer for different pumping and decay time-scales.}
    \label{fig:exc_no}
\end{figure}

As a last class of problems, let us move on to a system being simultaneously
pumped and drained of excitation from two different sites. Consider the same
excitonic dimer prepared in ground state as discussed in
subsection~\ref{sub:pump} which, in addition to being pumped from monomer 1, is
now also being drained from monomer 2 which are accounted for by the Lindblad
jump operators:
\begin{align}
    L_1^\text{pump} &= T_\text{pump}^{-1/2}\left(\dyad{eg}{gg} +\dyad{ee}{ge}\right) \\
    L_2^\text{drain} &= T_\text{drain}^{-1/2}\left(\dyad{gg}{ge} +\dyad{eg}{ee}\right)
\end{align}

We start by exploring the time-evolution of the excitation population, $\mathcal{E}_\alpha(t)=\Tr_\text{sys}\left[\rho_\text{sys}(t)\dyad{e_\alpha}\right]$, on the
monomers $\alpha$ in Fig.~\ref{fig:excitation_dimer} for three different combinations of
pumping and draining time-scales. Trivially, when the pumping rate is faster
than draining rate, then the rise of the excitation population in the system is
the largest. Notice that for each of these combinations, a steady-state is
reached in the system.

\begin{figure}
    \centering
    \hspace*{-0.5cm}
    \subfloat[Flow into $\ket{gg}$]{\includegraphics{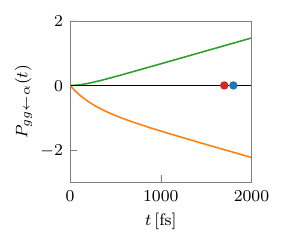}}
    \subfloat[Flow into $\ket{eg}$]{\includegraphics{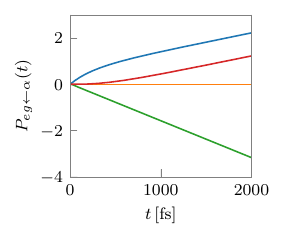}}

    \hspace*{-0.5cm}
    \subfloat[Flow into $\ket{ge}$]{\includegraphics{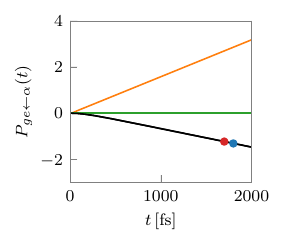}}
    \subfloat[Flow into $\ket{ee}$]{\includegraphics{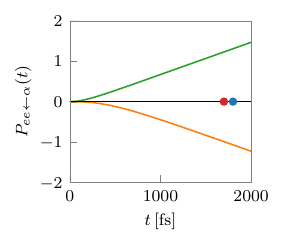}}
    
    \caption{State-to-state analysis of excitation flows into different diabatic states of the excitonic dimer when simultaneously pumped and drained with $T_\text{pump}=T_\text{drain}=\SI{300}{\fs}$ (Blue: $P_{*\leftarrow gg}(t)$, Orange: $P_{*\leftarrow eg}(t)$, Green: $P_{*\leftarrow ge}(t)$ and Red: $P_{*\leftarrow ee}(t)$. Black indicates multiple overlapping curves with legends marked as discs of corresponding colors.)}
    \label{fig:state_pump_drain}
\end{figure}

Before analyzing the dynamics further using the state-to-state method, let us
try to explore the steady state a bit more. We plot the total excitation ($\sum_\alpha\mathcal{E}_\alpha(t)$) at
long-times (after the steady-state has set in) in Fig.~\ref{fig:exc_no} as a function of the pumping and the
draining time-scales. Along the diagonal characterized by
$T_\text{pump}=T_\text{drain}$, one notices that the system has exactly a single excitation at
steady-state. When $T_\text{pump}>T_\text{drain}$, the excitation population
is less than one, and it is greater than one otherwise. Our preliminary explorations indicate that so long as the system is homogeneous (that is the monomers are identical) these values are independent of the type of vibrational bath and only dependent on the system description.

For making our state-to-state discussions concrete, let us pick the case of
$T_\text{pump}=T_\text{drain}=\SI{300}{\fs}$. For this case, the converged dynamics was recovered at $\Delta t=\SI{2}{\fs}$ and a memory time of $\tau_\text{mem}=\SI{60}{\fs}$. In
Fig.~\ref{fig:state_pump_drain}, we show the state-to-state transfers in the
diabatic basis. All the transfers to and from $\ket{gg}$ or $\ket{ee}$ are
Lindbladian in origin, and the Hamiltonian transports occur between the
$\ket{eg}$ and $\ket{ge}$ states. The interpretation is similar to the pure
pumping case and therefore, we skip it. One feature that is different from the
previous case and consequently, deserves mentioning  is the existence of certain
transport curves that asymptotically become straight lines, but with non-zero
gradients (eg. the Lindbladian transport into and from $\ket{gg}$). This means
that there is a continuous transport either into or from that state. This is
because the steady-states reached in these systems are dynamic, and a result of
balancing of the pumping and draining processes, both of which individually
proceed in their own ways.

\begin{figure}
    \centering
    \includegraphics{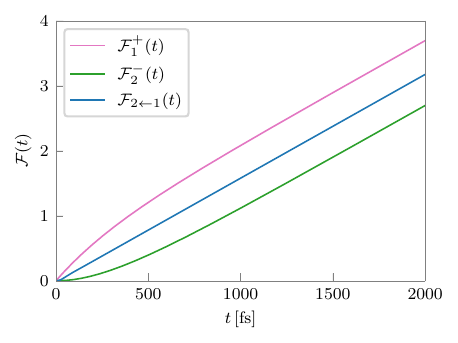}
    \caption{Total excitation flows $\mathcal{F}(t)$ into ($\mathcal{F}^{+}_{\alpha}(t)$), out of ($\mathcal{F}^{-}_{\alpha}(t)$) and between ($\mathcal{F}_{\alpha\leftarrow\beta}(t)$) monomers $\alpha$ (and $\beta$) of the excitonic dimer being pumped and drained simultaneously from monomers $1$ and $2$, respectively, with corresponding timescales $T_\text{pump}=T_\text{drain}=\SI{300}{\fs}$.}
    \label{fig:excitation_flow_dimer}
\end{figure}

\begin{figure}
    \centering
    \includegraphics{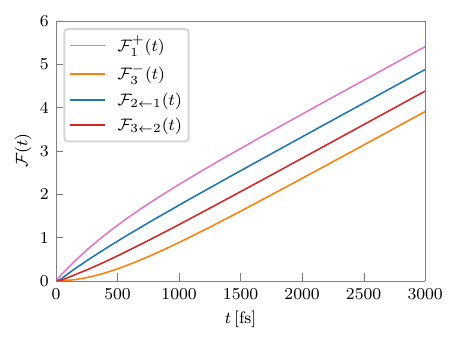}
    \caption{Total excitation flows $\mathcal{F}(t)$ into ($\mathcal{F}^{+}_{\alpha}(t)$), out of ($\mathcal{F}^{-}_{\alpha}(t)$) and between ($\mathcal{F}_{\alpha\leftarrow\beta}(t)$) monomers $\alpha$ (and $\beta$) of the excitonic trimer being pumped and drained simultaneously from monomers $1$ and $3$, respectively, with corresponding timescales $T_\text{pump}=T_\text{drain}=\SI{300}{\fs}$.}
    \label{fig:excitation_flow_trimer}
\end{figure}

At this stage, we switch to the state-to-state analysis defined in terms of the
monomeric excitation flows. Given that we are interested in spatial transport of
excitation across monomers, this is the more physically relevant basis. While
the diabatic state-to-state transfers (Fig.~\ref{fig:state_pump_drain}) are
directly measured, as mentioned previously (in Sec.~\ref{sub:pump}), the flow
between the monomers can easily be reconstructed from them. This is presented in
Fig.~\ref{fig:excitation_flow_dimer}. For our dimeric system, there are three
processes of interest --- (a) the Lindbladian pumping of monomer 1
($\mathcal{F}^+_1(t)$), (b) the Hamiltonian transport between monomers 1 and 2
($\mathcal{F}_{2\leftarrow 1}(t)$), and (c) the Lindbladian draining of monomer
2 ($\mathcal{F}^-_2(t)$). The total excitation content in the system at time $t$
is $\mathcal{F}_1^+(t) - \mathcal{F}_2^-(t)$. The buildup of excitation in the
system happens because of the time that it takes for the excitation to go from the
pumping site to the draining site. The lines, $\mathcal{F}^\pm(t)$, becoming parallel is a signature of
the steady-state setting in. Additionally, notice that because of the
simultaneous pumping and draining, there is a net current of excitonic
extraction from the second site. This is a feature that cannot be there for
aggregates with only draining sites.

\begin{figure}
    \centering
    \includegraphics{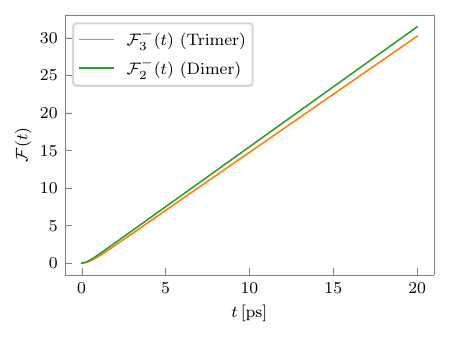}
    \caption{Excitation extraction as a function of time for the excitonic dimer and trimer with $T_\text{pump}=T_\text{drain}=\SI{300}{\fs}$.}
    \label{fig:excitation_flow_dimer_vs_trimer}
\end{figure}

As a final example, we study the excitonic trimer (defined with Eq.~\ref{eq:exciton_fullspace} for $N=3$) with pumping and
draining on the terminal sites (monomers $1$ and $3$, respectively). The flows in the monomeric picture are presented in
Fig.~\ref{fig:excitation_flow_trimer}. The same pattern emerges, except the
steady-state excitation content of the excitonic trimer $\lim_{t\to\infty}\mathcal{F}^+_1(t) - \mathcal{F}^-_3(t)$ is $1.5$ which is larger than 1.0 for the dimer.
We also notice that the only non-zero unmediated transports are between
consecutive monomers. There is no direct flow between monomer 1 and monomer 3.
This is because of the nearest neighbor couplings present in the system Hamiltonian, $\hat{H}_\text{sys}$. In
Fig.~\ref{fig:excitation_flow_dimer_vs_trimer} we show the extraction dynamics
of the exciton from the dimer and trimer. Notice that if one defines
the excitonic current as $I_\text{exc}(t) = \lim_{t\to\infty}
\frac{d\mathcal{F}_\text{drain}(t)}{dt}$, then it is clear that the dimer
provides a ``higher'' current ($I_\text{exc}^\text{dimer} =
\SI{1.597}{\per\ps}$) than the trimer
($I_\text{exc}^\text{trimer}=\SI{1.542}{\per\ps}$) even though both are made of
the same identical monomeric units. This size dependence and the other factors
affecting the excitonic current would be further explored in a future work.

\section{Conclusions}\label{sec:conclusions}

The state-to-state method~\cite{boseImpactSolventStatetoState2023,
boseImpactSpatialInhomogeneity2023} based on analysis of
fluxes~\cite{wuEfficientEnergyTransfer2012, boseNonequilibriumReactiveFlux2017,
daniQuantumStatetoStateRates2022} has proven to be quite capable of elucidating dynamical routes of transport in aggregates, shedding light on the many-body
effects of the environment as well. We had previously extended this framework to
non-Hermitian systems to enable calculation of transport efficiencies and
mechanisms in non-equilibrium processes involving a combination of thermal
environments and empirically defined
losses~\cite{sharmaImpactLossMechanisms2024}. In the current paper, we
investigate the possibilities of exploring the dynamics with greater granularity
when there are multiple pumps and drains affecting the system. We present a
generalization of the state-to-state transport analysis which incorporates
Lindbladian terms in addition to the effect of thermal solvents, going beyond our own
recently developed non-Hermitian state-to-state technique. A crucial aspect of such ``mixed'' simulations is that the empirical processes are treated under a Markovian approximation using the Lindblad master equation, while the solvent is treated using numerically exact path integrals capturing the non-Markovian memory effects. These empirical processes can either pump or drain the system, or be ones which
cause spin decoherence among many others. 

We started by validating our Lindblad state-to-state method against the non-Hermitian state-to-state analysis for lossy
systems. The results from both the methods are consistent. Then we demonstrated
cases involving pumping processes which only a Lindblad-like description can
handle. The Lindblad state-to-state method uncovered transport pathways in purely
pumped and simultaneously pumped and drained excitonic aggregates. For the
latter case, we showed the emergence of a steady-state current across the
aggregate. Surprisingly, this current is also a function of the aggregate size,
even for aggregates of identical monomers. Our Lindblad state-to-state method
provides a powerful framework with which one can start to explore these systems.
Given the richness of these systems, there are many other important parameters
to consider and explore. Future work will focus on these and try to deepen our
understanding of excitonic currents in transport systems.

One of the hallmarks of the family of state-to-state methods has been the
independence of the analysis from the actual method of simulation of the
dynamics. This feature is retained in the current work as well. While we have
used path integral methods to generate the
dynamics~\cite{strathearnEfficientNonMarkovianQuantum2018,
boseIncorporationEmpiricalGain2024} here, our method offers the flexibility to
use even semiclassical or perturbative methods. Moreover, using a variety of
Lindblad jump operators, one can incorporate processes beyond just pumping and
draining and study their effects. This generality makes the Lindblad state-to-state
transport analysis method extremely lucrative for unraveling the complexities of
the quantum transport in large aggregates.

\appendix

\section{Non-Hermitian Hamiltonians and Lindblad Jump Operators}\label{app:nh_vs_lindblad}

Consider the Lindblad master equation (Eq.~\ref{eq:lindbladME}), rewritten here for the sake of convenience:
\begin{align}
    \dot\rho(t) &= -\frac{i}{\hbar}\comm{\hat{H}}{\rho(t)} \nonumber\\
    &+ \sum_n \left(L_n\rho(t)L_n^\dag - \frac{1}{2}\acomm{L_n^\dag L_n}{\rho(t)}\right).\label{eq:lme}
\end{align}
The effect of the Lindbladians on the density matrix $\rho(t)$ above can be split into two parts~\cite{wiseman2009quantum}: the continuous non-unitary dissipation terms $\acomm{L_n^\dag L_n}{\rho(t)}$ and the quantum jump terms $L_n\rho(t) L_n^\dag$. The Lindblad master equation is overall trace-preserving and completely positive.

On introducing an effective non-Hermitian Hamiltonian of the form:
\begin{align}
    \hat{H}_\text{eff}=\hat{H}-i\sum_n L_n^\dag L_n /2\, ,\label{eq:nh_hamiltonian}
\end{align}
we note that the master equation in Eq.~\ref{eq:lme} can be rewritten in terms of $\hat{H}_\text{eff}$ as follows:
\begin{align}
    \dot\rho(t)&=-\frac{i}{\hbar}\left(\hat{H}_\text{eff}\rho(t)-\rho(t)\hat{H}^\dag_\text{eff}\right) \nonumber \\
    &+\sum_n L_n\rho(t)L_n^\dag \label{eq:efflme}
\end{align}
If one were to ignore the last term ($\sum_n L_n\rho(t)L_n^\dag$) in Eq.~\ref{eq:efflme}, the equation becomes the equation of motion for a generic non-Hermitian Hamiltonian. However, as evident, the time evolution no longer satisfies the property of being trace-conserving and completely positive.

This inability of a non-Hermitian Hamiltonian to conserve the trace of the density matrix brings forth several limitations. One such limitation is the fact that there is no commensurate rise in the population of the states into which the system decays. For example, in a two-level system with a non-Hermitian Hamiltonian describing the decay of excited state $\ket{e}$, the decrease in the population of $\ket{e}$ would not lead to any change in the population of the ground state $\ket{g}$.  As such, this would introduce spurious effects in absorption spectrum or any other observable requiring trace-conservation.

Another limitation of the non-Hermitian description is its inability to describe pump processes. If one were to use a non-Hermitian Hamiltonian with a positive imaginary part, say, $\hat{H}_\text{eff}=\hat{H}+i\Gamma/2$ for some diagonal and Hermitian operator $\Gamma$, in hopes of simulating a pump (instead of a dissipation/loss always introduced with a negative imaginary part), then one would get $\rho(t)\propto e^{\Gamma t}\rho(0)$ for a purely non-Hermitian time evolution. This leads to an exponential rise of the population of the pumped states, provided they start with a non-zero initial population, which is phenomenologically incorrect. Additionally, the states with zero initial population will never rise on being pumped in such a fashion.


\bibliography{library}
\end{document}